\documentclass[sigconf]{acmart}

\usepackage{booktabs} 
\usepackage{enumitem}
\usepackage{graphicx}
\usepackage{color}
\usepackage{cancel}
\usepackage[normalem]{ulem}
\newcommand\redsout{\bgroup\markoverwith{\textcolor{red}{\rule[0.5ex]{2pt}{0.4pt}}}\ULon}
\newcommand{\sob}[1]{\textcolor{blue}{#1}}
\usepackage{blindtext, graphicx}
\setcopyright{rightsretained}

\copyrightyear{2017}
\acmYear{2017}
\setcopyright{othergov}
\acmConference{ICFNDS '17}{July 19-20, 2017}{Cambridge, United Kingdom}
\acmPrice{15.00}
\acmDOI{10.1145/3102304.3102345}
\acmISBN{978-1-4503-4844-7/17/07}

\begin{document}
\title{Towards Information-Centric Networking (ICN) Naming for Internet of Things (IoT):}
\subtitle{The Case of Smart Campus}

\author{Sobia Arshad}
\authornote{Working as PhD full-time Researcher at UET Taxila, Pakistan}
\orcid{1234-5678-9012}
\orcid{0000-0002-9962-3536}
\affiliation{%
  \institution{Department of Computer Engineering}
  \streetaddress{University of Engineering \& Technology (UET)}
  \city{Taxila}
  \country{Pakistan} 
}
\email{sobia.arshad@uettaxila.edu.pk}

\author{Muhammad~Awais~Azam}
\affiliation{%
   \institution{Department of Computer Engineering}
  \streetaddress{University of Engineering \& Technology (UET)}
  \city{Taxila}
  \country{Pakistan} 
}
\email{awais.azam@uettaxila.edu.pk}

\author{Syed Hassan Ahmed}
  \affiliation{%
\institution{School of computer Science and Engineering}
  \streetaddress{Kyungpook National University}
  \city{Daegu}
  \country{Republic of Korea} 
  }
\email{s.h.ahmed@ieee.org}

\author{Prof. Jonathan~Loo}
\affiliation{
 \institution{School of Computing and Engineering}
  \streetaddress{University of West London}
  \city{London}
  \country{United Kingdom}
 }
\email{jonathan.loo@uwl.ac.uk}





\renewcommand{\shortauthors}{S. Arshad et al.}

\begin{abstract}
Information-Centric Networking (ICN) specifically Name Data Networking (NDN) is the name-base (content-base)  networking and takes named-contents as "first class citizen", being considered as the ideal candidate to form the Future Internet basis. NDN striking features like named-data self-secured contents, name-base-forwarding, in-network caching and mobility support suits the Internet of Things (IoT) environment, which aims to enable communication among smart devices and to combine all Internet-based smart applications under the one roof. With these aims, IoT put many research challenges regarding its network architecture as it should support heterogeneous devices and offer scalability. IoT may depend on the names and addresses of billions of the devices and should smartly manage the bulk of data produced every second. IoT application smart campus has gained a lot of attention in both industry and academia due to many reasons.
Therefore, to design NDN for IoT, a sophisticated naming scheme is needed to explore and it is the main motivation for this work. In this paper, we study NDN-IoT smart campus (in terms of connected devices and contents) and find that it lacks in a reasonable naming and addressing mechanism; and thus we propose NDN based Hybrid Naming Scheme (NDN-HNS) for IoT based Smart Campus (IoTSC).
\end{abstract}

%
%



\maketitle

\section{Introduction}

%
%
%
%


 

As smart connected devices usage prevalent now-a-days. And plethora of these smart connected devices is referred as Internet of Things (IoT) \cite{jabbar2016rest}-\cite{ahmad2016smart}-\cite{ahmad2016efficient}. These smart connected devices may include smart phones, tablets, smart gadgets like iWatch, smart glass, smart brush, smart AC and gained much attention from consumers as well as from both industry and research societies. For example, IoT European Research Cluster (IERC) aims to identify the challenges to transform the IoT concept into IoT reality. IERC involves other countries than Europe to address global connectivity \cite{ierc}. Another, ICN Research Group (ICNRG) is also trying to gather requirements to build IoT architecture using ICN \cite{icnrg}. But, with the advent of IoT, its efficient network architecture is still in its infancy.


 %

In the context of ICN, architectures are proposed like NDN, PURSUIT, SAIL, MobilityFirst, CONVERGENCE, COMET, Green ICN and C-DAX \cite{ICNsurvey}-\cite{AmadeoCampoloQuevedoEtAl2016}. ICN has many research areas to be explored yet and naming the contents is one of them. Among these ICN architectures, NDN is implemented as proof-of-concept and gaining a lot of attraction through research community \cite{ccnx-prominent-proof-of-concept}.

Hierarchical naming scheme is used to name the contents in NDN. This naming scheme provides long variable length names \cite{ICNsurvey}.  
\begin{figure}[!t]
\centering
\includegraphics[width=9cm]{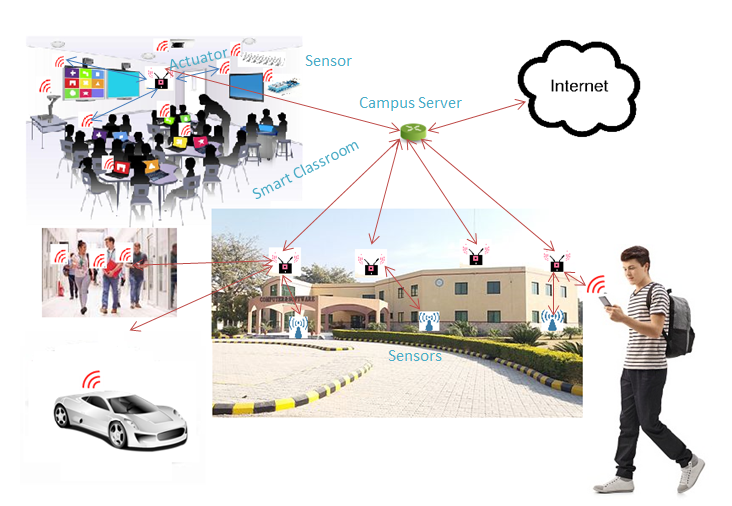}
\caption{Smart Campus, Computer Engineering Department (CPED), UET, Taxila.}
\label{smart-campus}
\end{figure}
On the other hand, in general, IoTs contents are ephemeral, short-lived, small, fresh, different priorities and from different locations. Therefore IoTs contents have special features and factors that need to be accommodated/ considered while naming any specific content. 

In this paper, we select \textbf{\textit{"Smart Campus"}} as an IoT usecase, as shown in Fig.~\ref{smart-campus}. We take smart campus because it involves almost all operations that any other IoT application does and research for smart campus is a never ending process as we can keep adding smart services to make it more smarter. At the same time, it is very emerging research topic from both academics and industry perspectives\cite{agarwal2016toward}-\cite{xu2016toward}. Moreover, 
in cases like smart home, IoTs scalibility may not be estimated appropriately due to limited number of nodes and data. 

Further, we state IoTs based Smart Campus (IoTSC) as a building or combination of many buildings that may utilize(s) many smart sensors (temperature, humidity, pressure, proximity and etc.,), actuators (AC, lights, fans, doors, windows, vehicles, mobile phones, alarm buzzers) and connecting devices (Ethernet, Wifi, Bluetooth) to provide anytime connectivity. CAmpus Server (CAS) is responsible to provide connectivity among all devices and monitor all activities as controlling agent as can be seen in Fig.~\ref{smart-campus}. NDN based IoTSC application can provide many services as energy management, security and privacy. 

Then, in this paper, we propose NDN-based Hybrid Naming Scheme (NDN-HNS) to name contents and devices considering best features from ICN naming schemes \cite{zhang2016uniform} for IoTs environment. Providing holistic naming scheme for NDN-IoT is the major contribution of this work. In this scheme, name is assigned to content in three components: IoT application prefix, hierarchical, flat-hash and attribute-based. Proposed naming scheme can provide scalability, security and addressing and naming to data contents and devices.        

Rest of the paper follows with some background, importance and NDN based research efforts of smart campus w.r.t IoT in section II. Proposed NDN-HNS and its important parts are described in section III 
and finally we conclude our so-far work in conclusions section.

\begin{figure*}[t]
\centering
\includegraphics[width=19cm]{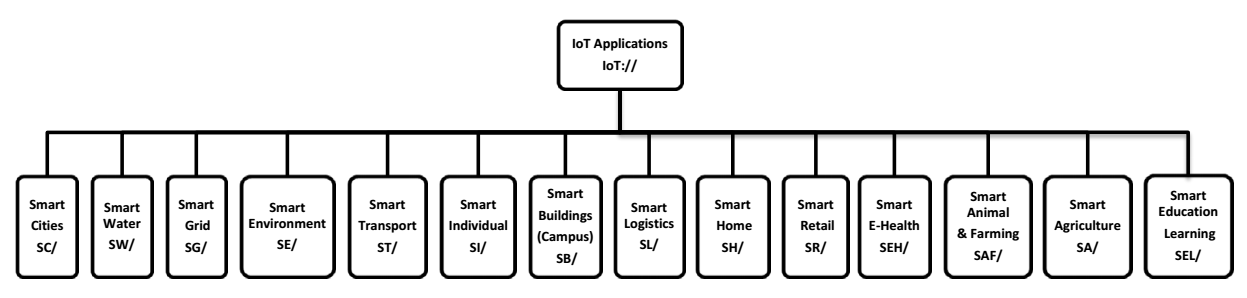}
\caption{IoT Applications Categorization.}
\label{iot-app}
\end{figure*}

\section{IoT based Smart Campus and NDN Related Research Efforts}
This section firstly, presents an overview of smart campus importance in IoT from both academics and industrial perspectives. Further we describe why TCP/IP is less suitable for IoT. Then we discuss some research efforts in which NDN is considered for IoT and smart campus.

 Google funded and started GIoTTO program at CMU with the collaborators from other universities as well to transform IoT into reality and under GIoTTO they build lab to convert CMU into smart campus\cite{agarwal2016toward}. They designed GIoTTO open source stack to provide support for heterogeneous platforms. Moreover, Google invested in NEST Labs for IoT, which aim to build smart home \cite{xu2016toward}. On the other hand, Huawei used IPv4 and IPv6 to provide smart campus services to Huawei Hubei University of Technology \cite{huawei-smart-campus}. They aim to provide smart agile campus, cloud data center, disaster management and recovery mode, easy access to library and safety of campus through video surveillance. In addition, Pakistani government has announced that thirteen universities are being converted to smart campus by the end of year 2017 \cite{hec-smart-campus}. As a first step, Pakistani government has decided to provide students and faculty with free wifi everywhere inside the campus. Next step is to utilize campus resources to monitor energy usage to control and optimize its consumption. 

Major activities/services that can help to build IoTSC may involve optimization and control of energy usage  \cite{lazaroiu2016energy}, security of faculty and students vehicles, security and privacy of data of both faculty and students, easy access of library resources, smart cafeteria ordering and students behavior analysis. Other advance services may include disaster management, hostel allotment and mess management system, smart attendance system and smart time table management system. Mostly services can be achieved through simple operation of connectivity among smart devices i.e., both sensors and actuators. While other services may be achieved through smart mobile applications development, cloud computing, artificial intelligence algorithms and neural networks \cite{adamko2014intelligent}-\cite{guo2015research}. 

While we argue that employing TCP/IP architecture for IoTSC can burdensome constraint-oriented devices in many ways, for example: 

i) Data produced by billions tiny IoT devices is huge in amount and needs better management to use this data for analytics applications,

ii) Mostly IoT devices (especially wireless sensors) come up in small memory due to the processor used and the circuit board design. Small memory makes data unavailable frequently, further, there is no caching system in IP-based solutions which also raises data unavailability issue,
 
iii) IoT can definitely have devices that differ in specifications (differ in memory, power, communication range and processing power) and connection technologies (UWB, RFID, Zigbee, Ethernet, Bluetooth, Wi-Fi, Wi-MAX, MANETs, Cellular Networks, Radars and Satellite Networks (Although last two technologies are not meant for smart campus but may involve in IoT))\cite{van2016flexible},

iv) Scanning and inspection of transmitted data need complex and costly methods like Deep Packet Inspection (DPI),

v) Separate patches like IP-Sec protocols are employed to provide appropriate security and

vi) TCP/IP provides mobility support in complex way where multiple additional registrations (during mobile devices hand-off) are required to support mobile devices. Security requirements and mobile devices further complicates its  applicability for IoTs. 

Therefore, heavy TCP/IP architecture, either in native or in overlay manner, is less suitable for IoTSC.

In the meantime, many research efforts suggest ICN, specifically NDN for IoT smart environment. NDN works on publish/subscribe model and offer name base networking through two messages namely Interest Message and Data Message, and  three data structures namely Pending Interest Table (PIT), Forwarding Information Base (FIB) and Content Store (CS) \cite{shang2016named}. Subscriber need to subscribe the network to get data of its interest that can be requested through Interest Message. Publisher advertises it's produced data towards nearest content routers (CRs). CRs directly forwards data through Data Message if it is stored in CS on finding the prefix match with content name mentioned in Interest Message and Interest Message is dropped.  Else CR looks for entry of forwarding path in FIB.  CR performs prefix match in FIB to find the next CR(s) list to forward Interest Message and after forwarding Interest message towards next CR, it places the entry incoming interface in PIT.  CR adds incoming interface of this entry if a match of this entry is already available in PIT and Interest Message is dropped. When a DATA Message received by any CR it caches this data in its CS though 'Cache each everything (CE2)' caching scheme. Then CR finds prefix (es) for appropriate interface (s) in PIT and upon a successful matching it routes this data towards that (or those) interfaces (s). 
NDN has been implemented for smart home in both \cite{AmadeoCampoloIeraEtAl2015}-\cite{ahmed2016named}. Both of these works discuss naive NDN hierarchical naming scheme for smart home. Further \cite{de2016named} implement NDN for lightning the home and found NDN more useful when traffic is local. They also implement NDN basic naming scheme for naming the data. Smart campus lightning system is implemented through NDN in \cite{de2016implementation}. They also use basic NDN hierarchical naming scheme to name the contents.

\section{Hybrid ICN-based Naming Scheme for IoTs}
\sob{This section is presented with threefold purposes: in the first Subsection IoT application categorization is presented and in the Subsection section hybrid naming concept is described. In the third Subsection, naming scheme components are discussed.}
\subsection{IoT Application Categorization}
We use \cite{iotapps} to categorize IoT applications and show in Fig.~\ref{iot-app}  . They listed 12 major categories and we updated (and somehow modified) this list by adding two more categories named as smart education learning and smart buildings. 
\subsubsection{Smart Cities} 
This application can include structure health monitoring (SHM), roads and their lighting management, traffic congestion management, parking lots management, waste management and noise control to build smart city. 
\subsubsection{Smart Water}
This scenario can incorporates water monitoring through pollution levels in the seas and rivers, water level monitoring for flood control and chemicals proportion monitoring for purity water.   
\subsubsection{Smart Grid}
This application can handle smart metering, electricity usage control and automatic billing.
\subsubsection{Smart Environment}
This can incorporate earthquake detection, snow level and landslide monitoring and management, fire detection and control in the towns and in the forests to make environment safe and sound to provide better life.
\subsubsection{Smart Transportation}
Efficient transfer of wide range of commodities can be included in this scenario. Vehicles carrying a commodity can be tagged and located via GPRS software. 
\subsubsection{Smart Individual}
As mentioned in \cite{smart-individual} IoT through smart wearable can be simple IoMe, to control your daily intake of calories, to share health measurements with your doctor and to motivate you to be more active in sports. 
\subsubsection{Smart Buildings}
In \cite{smart-building}, smart building can incorporate chillers, air handlers, automatic parking, energy consumption control and temperature and humidity monitoring.  
\subsubsection{Smart Logistics}
\sob{This can incorporate }planning, policy and infrastructure \cite{van2012smart} to manage route control, easy detection of  an item in supply chain, quality control of shipments to ensure safety  and storage compatibility management. 
\subsubsection{Smart Home} 
Security system to detect and monitor entrance of non-authorize persons, automatic control (locking and opening) of house doors and windows, consumption control of water, gas and electricity, remote control and monitor of house appliances can be combined to build smart home.  
\subsubsection{ Smart Retail}
This can include applications like smart shopping and payment management, product store management, payment management for parks, roads and gyms. 
\subsubsection{ Smart E-Health}
Check and control of sugar level and blood pressure, measurement of daily calories intake, assistance of elderly Alzheimer patients  and proper medication (in case of emergency) through remote help from doctors can incorporate in providing smart-e-health.
\subsubsection{ Smart Animal \& Farming}
Quality control and improvement in survival rate of offspring, betterment in ventilation conditions, animal tracking and control of required conditions to ensure high quality crops can be combined to form smart animal farming. 
\subsubsection{ Smart Agriculture}
This can monitor and control soil moisture level, humidity, temperature and wind changes to maximize product (fruits and vegetables) quality and quantity.
\subsubsection{ Smart Education Learning} 
This scenario can include smart campus services like notification for exam results, evaluation of assignments and exams, and important announcements (about class timings and venues). Moreover, it can include energy management in campus.    
\begin{figure*}[!t]
\centering
\includegraphics[width=19cm]{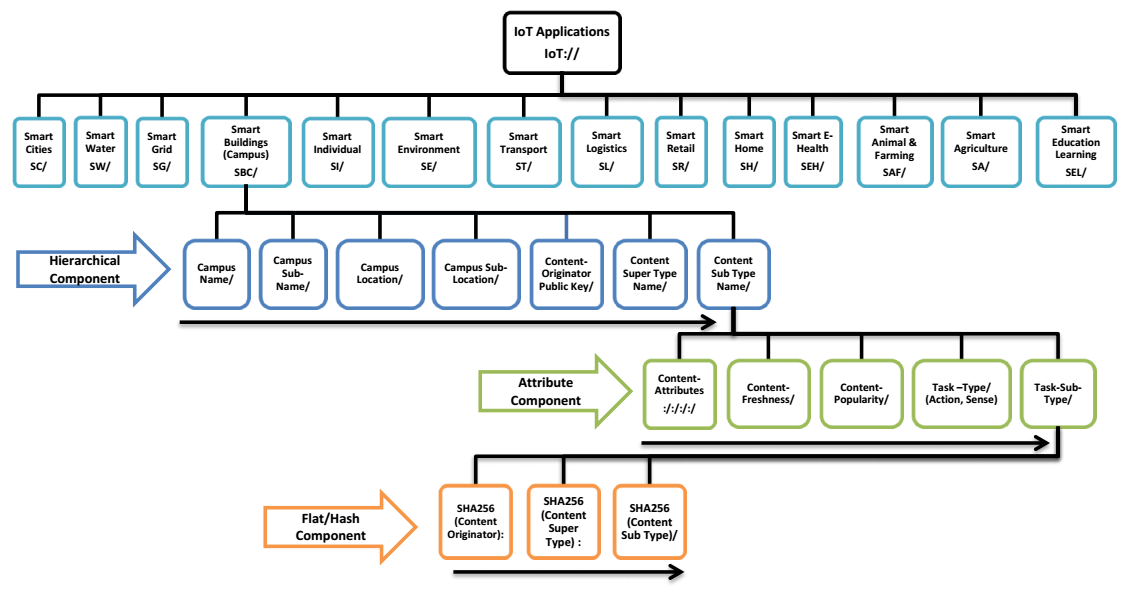}
\caption{IoTs applications naming and resolution. Step 1: Hierarchical Part and resolution. Step 2: Attribute Part and resolution. Step 3: Flat/Hash Part and resolution.}
\label{iot-naming-scheme}
\end{figure*}
\subsection{Proposed NDN-HNS for IoTSC}
When an IoT application user (for example IoTSC user) requests for a content, NDN Interest message is forwarded from the requesting node to nearest node (or intermediate node). This node further forwards Interest message to the node containing the content. Any node that originates data or caches data can reply DATA message. NDN model follows hierarchical naming scheme to name IoT contents. Hierarchical names are long and describe detail of content separating through /. NDN naming scheme is not designed for IoT applications as these application put many new constraints to name the contents. \sob{To fulfill IoT application specific requirements we propose a  hybrid }and holistic naming scheme which contains multiple parts to name any content. \sob{These parts or components are separated through the symbol ':', that} indicates the end of current part and start of the next part. Moreover, :/ is used to specify multiple attributes of the content in attributes component and multiple sub-parts of flat component. While portions of these parts are specified through /. 
 
 \subsection{\sob{NDN-HNS Components}}
Our hybrid naming approach contains following three parts. These parts are aggregated with the aim to provide scalable, secure and easy-to-manipulate naming scheme. But before these name parts begin, user has to specify primary root prefix as \textit{IoT://SBC} for \textit{Smart Building (Campus) IoT application} as it can be seen in Fig.~\ref{iot-naming-scheme}. Then name follows with hierarchical component as secondary hierarchical root prefix, attribute component and flat self-certifying component. 
 \subsubsection{Hierarchical Component (HC) or Secondary Root Prefix (SRP)}
This part names a content in the same way as NDN assigns name. Hierarchical component (HC) combines campus information, content originating node ID along with content information. 
However, campus location information has nothing to do with the physical location of campus instead it has just to represent name of location and sub-location of campus. 
HC's content information (both super type and sub type) is added to identify content main category and sub category. So that, HC can provide easy name management and aggregation, simple and easy search, and optimized routing tables 
by including above mentioned information about campus and content. HC part's information is described as follows:  
\begin{itemize}
\item{\textbf{Campus Name/Campus Sub-Name/:}} This refers to the name of campus, for example, in our scenario, UET Taxila/CPED values refer to the Computer Engineering Department (CPED) of University of Engineering and Technology (UET), Taxila.
\item{\textbf{Campus Location/Campus Sub-Location/:}} represent campus's country and city. In other words, where this campus is situated (or about which campus content is required), e.g., Pakistan/Taxila represents CPED is located in Taxila city of Pakistan.
\item{\textbf{Content Originator ID/Public Key:}} This may involve student's registration number/faculty's employee name (or/and number),  or sensor/actuator ID in case of devices e.g., 14F-UET-PhD-CP-43 represents registration number of Computer (CP) student and can refer to her/his cell phone device.
\item{\textbf{Content Super-Type Name/Content Sub-Type Name:}} represents the content's super type from  \textit{text, image, video etc.} and sub type from \textit{.word, .txt, .ppt etc.} for text, \textit{.jpg, .png etc.} for image and \textit{.avi, .mp4 etc.} for videos. For instance, \textit{Timetable-14CP/.xls} refers to the text file of Timetable-14CP in .xls (and not in any other format e.g., .word, .txt, .ppt). This ensures further the exact type of the required content. In addition, by this, required content can be searched easily.
  \end{itemize} 
Moreover, HC's sub parts can be clearly seen in the second line (in ink-blue color) of Fig.~\ref{iot-naming-scheme}.

\subsubsection{Attributes Component (AC)}
This component describes the details about a specific content. Attributes include content properties and the task type. 
AC's both task type and task sub-type specify type of communication that either sensing operation information is required or it will trigger/control some action. Here one thing is important to note that, inherent NDN supports only pull type communications. But with our naming scheme, push type communication will also be carried. Details of AC's sub-parts in following text:
\begin{itemize}
\item{\textbf{Content-Attributes:/:/:/:}} holds the values of multiple attributes for the desired content to signify it further, \sob{for example, \textit{14/01-Jan/13:30/1/} refers to }the extracted content's year-session (2014), date (1st January), time (13:30) and version (01) according to attributes  \textit{session/date/time/ver/}. 
\item{\textbf{Content-Freshness/:}} tells about content's life. One can specify about content's freshness in this, may be, by adding the time-stamp of content's generation time. IoTSC subscriber can get most recent and updated value by specifying this value as zero. And most old value can be get by specifying as one. This feature is added especially for sensors that provides some measurements ( e.g., temperature, humidity, etc.) to enable pure IoT functionality like getting the most recent value of temperature in campus. This can further help in content caching decision, which is currently beyond the scope of this work.  
\item{\textbf{Content-Popularity/:}} specify the popularity of the content. For instance, if a exam result is declared, then every student must be interested to get it. Contents can be easily be accessed (and provided) from near-by devices (even from friend's cell phone). Moreover, course video lectures marked as popular can be accessed more quickly. Its a simple counter that increments itself on every request. Moreover, other complex methods (e.g., Zipf's Distribution) can be explored to calculate popularity that helps in caching decision.  
\item{\textbf{Task-Type/Task-Sub-Type:}} holds the values about what is going to happen. Both \textit{Task-Type/Task-Sub-Type} specify which operation will happen. Whether it is some action with the values \textit{action/Turn-Light:(ON or OFF)} or sensing operation with the values \textit{sense/(Temperature or Humidity)}.  
  \end{itemize} 
Moreover, AC's sub parts can be clearly seen in the third line (in light green color) of Fig.~\ref{iot-naming-scheme}. 
\subsubsection{Flat Component (FC)}
This part is included to provide secure, signed and self-certified contents names. FC can hold the hashed value of either Content Originator ID or Content Super-Type Name or Content Sub-Type Name
\begin{itemize}
\item{\textbf{SHA256(Content Originator):/ SHA256 (Content Super-Type):/ SHA256(Content Sub-Type):/ :}} embeds hashed encrypted values of content originator and of content's name. For example,968cbab1de...:/e95e2bf0247.../0ac8b624229a...:/ represents SHA256s for 14F-UET-PhD-CP-43/Timetable-14CP/.xls. 
\end{itemize} 
Moreover, FC's sub parts can be clearly seen in the last line (in Orange color) of Fig.~\ref{iot-naming-scheme}. However, one can argue why to use SHA256, as it creates 32-bytes data even for a small content like \textit{.xls}. Therefore, one may use Base64 format to create FC. Study and selection for suitable creation methods for FC are part of future work. 

\section{Conclusion and Future Work}
We study ICN naming schemes for IoTs smart environment like smart home and smart campus. We proposed a sophisticated ICN (specifically NDN) based naming scheme for IoTSC. Moreover, we provide updated categorization of IoT applications. Proposed NDN-HNS operates in three important steps to assign a name to any content. As a first step, HC holds value about campus, content originator/requester and content itself. In second step AC contributes content's attributes, freshness, popularity and about task type. Security and authentication of the content is added by third component i.e., FC. 
We aim to evaluate proposed naming scheme for names aggregation and better request satisfaction rate for IoTSC. Moreover, we aim to provide (another mode of naming) simple and easy naming scheme for smaller contents. 
These are the key questions for the follow-up work.
\section*{Acknowledgment}
The authors would like to thank UET Taxila for PhD Fulltime Scholarship Program and to the faculty of computer engineering department for providing the appropriate research environment.

\bibliographystyle{ACM-Reference-Format}
\bibliography{iot-access}

\end{document}